# CN-LBP: Complex Networks-based Local Binary Patterns for Texture Classification

Zhengrui Huang

*Abstract*—To overcome the limitations of original local binary patterns (LBP), this article proposes a new texture descriptor aided by complex networks (CN) and LBP, named CN-LBP. Specifically, we first abstract a texture image (TI) as directed graphs over different bands with the help of pixel distance, intensity, and gradient (magnitude and angle). Second, several CN-based feature measurements (including clustering coefficient, in-degree centrality, out-degree centrality, and eigenvector centrality) are selected to further decipher the texture features, which generates four feature images that can retain the image information as much as possible. Third, given the original TIs, gradient images (GI), and generated feature images, we can obtain the discriminative representation of texture images based on uniform LBP (ULBP). Finally, the feature vector is obtained by jointly calculating and concatenating the spatial histograms. In contrast to original LBP, the proposed texture descriptor contains more detailed image information, and shows resistance to imaging and noise. Experiment results on four datasets demonstrate that the proposed texture descriptor can significantly improve the classification accuracies compared with the state-of-the-art LBP-based variants and deep learning-based methods.

## I. INTRODUCTION

RECENT advancements in pattern recognition (PR) provide many effective solutions to some applications, including ground object detection and classification [1], face recognition for industrial internet of things (IoT) [2], biomedical image retrieval [3], visual information sonification [4], cloud-based proactive caching [5], etc. In practice, the most widely studied part of PR is texture classification, because most of paradigms are developed based on it [6]. For accurate texture classification, a large number of approaches were summarized in [7], such as statistical approaches (e.g., gray level co-occurrence matrix (GLCM) [8], histogram of gradient (HoG) [9], local binary patterns (LBP) [10], and some LBP-based variants [11]-[12]), transform-based approaches (e.g., steerable filters [13], Gabor wavelet-assisted decomposition [14], and wedge filters [15]), model-based approaches (e.g., complex networks (CN) [16], gravitational models [12], and fractal dimension (FD) [17]), learning-based approaches (e.g., convolutional neural networks (CNN) [18], long-short term memory (LSTM) [19], and generative adversarial networks (GAN) [20]), etc.

Compared with other approaches, statistic-based descriptors (LBP) are used most widely. Original LBP was first proposed for rotation invariant texture classification in [10]. Based on the idea of local encoding patterns, some efficient LBP-based variants were introduced. For instance, in [21], Zhang *et al.* proposed a high-order information-based local derivative pattern (LDP) for face recognition and classification. To reduce the sensitivity of imaging conditions, a local energy pattern (LEP)-based descriptor was proposed in [22]. Following [22], a local directional number (LDN)-based descriptor was designed to distinguish similar structural patterns [23]. To reduce the impact of illumination and pose, Ding *et al.* introduced a dual-cross pattern (DCP) for facial expression detection [24]. Moreover, a local bit-plane decoded pattern (LBDP) was proposed for image matching and retrieval [3], a dominant rotated LBP (DRLBP) was studied to capture complimentary information [25], and Chakraborty *et al.* proposed a descriptor named centre symmetric quadruple pattern (CSQP) by encoding larger neighbourhood [26]. Other potential methods included local directional ternary pattern (LDTP) [27], local concave-and-convex micro-structure pattern (LCCMSP) [28], local neighbourhood difference pattern (LNDP) [29], etc. However, it is worth noting that most texture descriptors just focus on image texture information, but ignore the information of spatial-relationship, energy, entropy, or topology [30].

Motivated by the above, we comprehensively study in this article a texture descriptor by combining the characteristics of CN and LBP, namely, CN-based LBP (CN-LBP). Compared with the works in [10], and [21]-[30], we consider the practical case that CN mapping can effectively decipher the spatial-relationship between each pair of pixels and the distribution of pixels, and LBP encoding is rotation invariant and can reduce the impact of illumination. Specifically, we first map a texture image (TI) as directed graphs over different bands (RGB) based on pixel distance, intensity, and gradient, where the gradient information contains magnitude and angle. Second, the CN-based feature measurements (including *clustering coefficient* (CC), *in-degree centrality* (IDC), *out-degree centrality* (ODC), and *eigenvector centrality* (EC)) are selected to represent the characteristics (unique texture features) of the texture image. Next, we apply the uniform LBP (ULBP) on the TIs, gradient images (GI), and four feature images, and then obtain the final feature vector by jointly calculating and concatenating the spatial histograms, which is used for texture classification.

The remainder of this article is organized as follows. In

Zhengrui Huang and Chongcheng Chen are with the Academy of Digital China (Fujian), Fuzhou University, Fuzhou 350108, China, and the Key Laboratory of Spatial Data Mining & Information Sharing of Ministry Education, Fuzhou University, Fuzhou 350108, China.

Section II, we introduce the proposed texture descriptor. The experimental results and analyses are shown in Section III, and the conclusions are drawn in Section IV.

*Notations*: Scalars are denoted by italic letters, and vectors and matrices are denoted by bold-face lower-case and bold-face upper-case letters, respectively. For a real $x$, $\mathbb{I}(x)$ stands for the indicator function that equals 1 if $x > 0$, and 0 otherwise. For a set $\mathcal{X}$, $|\mathcal{X}|$ denotes its cardinality.

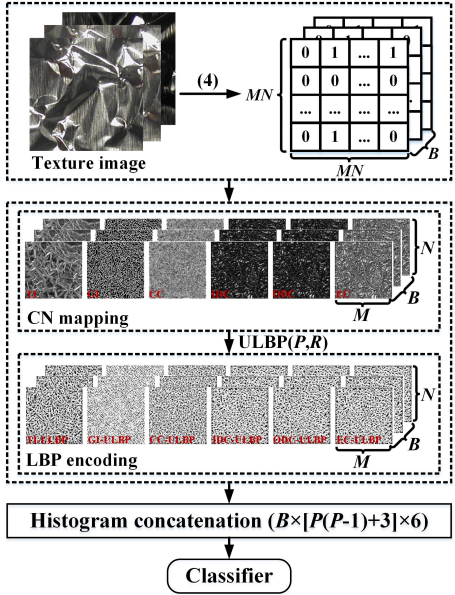

Fig. 1. Overview of CN-LBP.

## II. Proposed Method

In this section, we introduce CN mapping and LBP encoding respectively, as shown in Fig. 1.

### A. CN Mapping

Traditional CN model is actually a symmetric matrix and can be expressed as a graph, denoted by $\mathbf{G}_{M \times N \times B} = (\mathbf{V}, \mathbf{L})$, where $M \times N$ is the number of pixels over a band, $\mathbf{V}$ is the matrix of pixels, and $\mathbf{L}$ is the adjacent matrix, whose entry $l_{i,j}$ equals 1 if pixel $i$ links with pixel $j$, and 0 otherwise. Especially, for an undirected graph, $l_{i,j} = l_{j,i}$, and the degree of pixel $i$ denotes the number of edges connected with it [31]:

$$k_i = \sum_j l_{i,j} = \sum_j l_{j,i} \quad (1)$$

Since a CN model has unique features that are directly and indirectly developed based on (1). Thereby, by extracting these features from CN models, we can use them to discriminate differences between any two CN models, i.e., classify different texture images.

Given a texture image $\mathbf{I}_{M \times N \times B}$, the distance between pixel $i$ and pixel $j$ is defined as follows:

$$d_{i,j} = \sqrt{(x_i - x_j)^2 + (y_i - y_j)^2} \quad (2)$$

where $(x_i, y_i)$ and $(x_j, y_j)$ denote the image coordinates of pixel $i$ and pixel $j$, respectively, and the weight of $l_{i,j}$ is given by [32]:

$$w_{i,j} = \begin{cases} \dfrac{d_{i,j}^2 + q^2 |I_i - I_j|/L}{2q^2}, & \text{if } d_{i,j} \leq q \\ 0, & \text{otherwise} \end{cases} \quad (3)$$

where $q$ is the threshold of searching radius, $I$ is the pixel value, and $L$ is the maximum of gray scales.

Since traditional CN mapping only depends on (2)-(3), and does not take into account the gradient-related information. To make the CN mapping more robust, we introduce the gradient-based constraint, so $l_{i,j}$ can be redefined as follows:

$$l_{i,j} = \begin{cases} 1, & \text{if } w_{i,j} \leq r \text{ and } g_{i,j} \leq s \text{ and } \theta_{i,j} \leq t \\ 0, & \text{otherwise} \end{cases} \quad (4)$$

where $r$, $s$, and $t$ are the thresholds of similarity, gradient difference, and angle difference, respectively, and $g_{i,j}$ can be calculated by Sobel kernel-based convolution operation:

$$g_{i,j} = \sqrt{(\mathbf{M}_1 * \mathbf{I}_i)^2 + (\mathbf{M}_2 * \mathbf{I}_i)^2} - \sqrt{(\mathbf{M}_1 * \mathbf{I}_j)^2 + (\mathbf{M}_2 * \mathbf{I}_j)^2} \quad (5)$$

where $\mathbf{I}_i$ and $\mathbf{I}_j$ are the 3×3 matrices around pixel $i$ and pixel $j$, respectively, $\mathbf{M}_1 = \begin{bmatrix} -1 & 0 & 1 \\ -2 & 0 & 2 \\ -1 & 0 & 1 \end{bmatrix}$, $\mathbf{M}_2 = \begin{bmatrix} 1 & 2 & 1 \\ 0 & 0 & 0 \\ -1 & -2 & -1 \end{bmatrix}$, and $\theta_{i,j}$ is given by [33]:

$$\theta_{i,j} = \arctan\left(\frac{\mathbf{M}_2 * \mathbf{I}_i}{\mathbf{M}_1 * \mathbf{I}_i}\right) - \arctan\left(\frac{\mathbf{M}_2 * \mathbf{I}_j}{\mathbf{M}_1 * \mathbf{I}_j}\right) \quad (6)$$

Following (1)-(6), $\mathbf{I}_{M \times N \times B}$ can be mapped as a directed graph, and the GI of $\mathbf{I}_{M \times N \times B}$ is also obtained by (5). It is worth noting that $\mathbf{L}$ is not a symmetric matrix, i.e., $l_{i,j} \neq l_{j,i}$, so we need to traverse all image pixels, and the degree of pixel $i$ consists of in-degree and out-degree, namely, $k_i = k_i^{\text{in}} + k_i^{\text{out}}$:

$$\begin{cases} k_i^{\text{in}} = \sum_{j \in \mathcal{I}} l_{i,j} \\ k_i^{\text{out}} = \sum_{j \in \mathcal{I}} l_{j,i} \end{cases} \quad (7)$$

where $\mathcal{I} = \{0, \ldots MN - 1\}$.

Based on the adjacent matrix $\mathbf{L}$ derived from (4), four CN-based feature measurements are selected to describe the features of texture images, including CC, IDC, ODC, and EC.

Specifically, CC describes the spatial distribution of classes and is given by [34]:

$$\text{CC}(i) = \dfrac{(1/2) \sum_{j \in \mathcal{I}} \sum_{k \in \mathcal{I} \setminus \{i,j\}} (l_{i,j} + l_{j,i})^{1/3} (l_{i,k} + l_{k,i})^{1/3} (l_{j,k} + l_{k,j})^{1/3}}{\left(\sum_{j \in \mathcal{I}} l_{i,j} + \sum_{j \in \mathcal{I}} l_{j,i}\right)\left(\sum_{j \in \mathcal{I}} l_{i,j} + \sum_{j \in \mathcal{I}} l_{j,i} - 1\right) - 2\sum_{j \in \mathcal{I}} l_{i,j} l_{j,i}} \quad (8)$$

where $\text{CC} \in [0,1]$, and the pixels with higher (8) are clustered into the same community that has higher inter-group distances and lower intra-group distances.

For IDC and ODC, they can be expressed as follows based on (7):

$$\begin{cases} \text{IDC}(i) = \dfrac{k_i^{\text{in}}}{|\mathcal{I}|-1} \\ \text{ODC}(i) = \dfrac{k_i^{\text{out}}}{|\mathcal{I}|-1} \end{cases} \quad (9)$$

And EC of pixel $i$ can be formulated as [35]:

$$\text{EC}(i) = \lambda \sum_{j \in \mathcal{I}} l_{i,j} u_j \quad (10)$$

where $\lambda$ is the reciprocal of the maximum eigenvalue of $\mathbf{L}$, and $u_j$ is the element of the eigenvector derived from $\lambda^{-1}$.

### B. LBP Encoding

In this section, we encode the TIs, GIs, and feature images derived from (1)-(10), and jointly calculate and concatenate the spatial histograms to obtain the discriminative representation of texture images.

The LBP delineates the differences between a central pixel $c$ and its nearby $P$ pixels with a radius $R$:

$$\text{LBP}_{P,R}(I_c) = \sum_{p=0}^{P-1} \mathbb{I}(I_p - I_c) 2^p \quad (11)$$

where $I_c$ and $I_p$ denote the pixel values of pixel $c$ and pixel $p$ around pixel $c$, respectively. To reduce the impact of rotation and illumination on texture images, a uniform pattern of LBP was introduced [10]:

$$\text{ULBP}_{P,R}(I_c) = \begin{cases} \sum_{p=0}^{P-1} \mathbb{I}(I_p - I_c) 2^p, & \text{if } \mathbb{U}(\text{LBP}_{P,R}(I_C)) \leq 2 \\ P + 1, & \text{otherwise} \end{cases} \quad (12)$$

where $\mathbb{U}(\cdot)$ is the uniform pattern function:

$$\mathbb{U}(\text{LBP}_{P,R}(I_c)) = |\mathbb{I}(I_{P-1} - I_c) - \mathbb{I}(I_0 - I_c)| + \sum_{p=1}^{P-1} |\mathbb{I}(I_p - I_c) - \mathbb{I}(I_{p-1} - I_c)| \quad (13)$$

where the number of patterns decreases from $2^P$ to $P(P-1)+3$, which significantly reduces the dimension of feature vector.

Following (8)-(10), four feature images are generated. To make the proposed texture descriptor rotation invariant and reduce the dimension of feature vector, we respectively apply ULBP on TIs, GIs, and four feature images, and obtain six ULBP-based feature maps: TI-$\text{ULBP}_{P,R}^{q,r,s,t}$, GI-$\text{ULBP}_{P,R}^{q,r,s,t}$, CC-$\text{ULBP}_{P,R}^{q,r,s,t}$, IDC-$\text{ULBP}_{P,R}^{q,r,s,t}$, ODC-$\text{ULBP}_{P,R}^{q,r,s,t}$, and EC-$\text{ULBP}_{P,R}^{q,r,s,t}$, as shown in Fig. 1. In this article, we let $q=3$, $r=0.315$, $s=5$, and $t=45$. However, $P$ and $R$ are controllable, where $P \in \mathcal{P} = \{8, 16, 24\}$ and $R \in \mathcal{R} = \{1, 2, 3\}$, and we jointly calculate and concatenate the spatial histograms to form the final feature vector of CN-LBP:

$$\mathbf{f}_{\text{CN-LBP}} = \begin{bmatrix} \mathcal{H}_{\text{TI-ULBP}_{P,R}^{q,r,s,t}}, \mathcal{H}_{\text{GI-ULBP}_{P,R}^{q,r,s,t}}, \mathcal{H}_{\text{CC-ULBP}_{P,R}^{q,r,s,t}}, \\ \mathcal{H}_{\text{IDC-ULBP}_{P,R}^{q,r,s,t}}, \mathcal{H}_{\text{ODC-ULBP}_{P,R}^{q,r,s,t}}, \mathcal{H}_{\text{EC-ULBP}_{P,R}^{q,r,s,t}} \end{bmatrix} \quad (14)$$

where $\mathcal{H}$ is the spatial histogram, $|\mathcal{H}| = B \times [P(P-1)+3]$, and the combinations of $(P, R)$ include $(8,1)$, $(16,2)$, and $(24,3)$.

Different from the previous works [21]-[30], our proposed texture descriptor not only contains more image information (texture, structure, spatial-relationship, etc), but also shows strong resistance to rotation, illumination, and imaging.

### III. EXPERIMENTS AND ANALYSES

#### A. Datasets Description and Parameter Settings

In our experiments, we select four open source datasets to verify the effectiveness of our proposed method, including CUReT [36], KTH-TIPS2-b [37], Outex-TC-00013 [38], and SIRI-WHU [39]:

- CUReT: The dataset has 61 classes, and each class has 205 images (640×480 pixels).
- KTH-TIPS2-b: The dataset contains 11 categories, and each category has four files that include 4×108 images with a size of 200×200 pixels.
- Outex-TC-00013: The dataset contains 1360 images that are divided into 68 classes, an the image size is 128×128 pixels.
- SIRI-WHU: The dataset includes 12 categories, and the number of samples is 200 (200×200 pixels).

Before training classifiers, we reshape texture images in each dataset as $128 \times 128$ pixels and randomly split them into two parts (test size is 0.3), and train three types of classifiers:

- Support vector machine (SVM): The kernel functions are linear function (SVM-L) and radial basis function (SVM-RBF), and the penalty coefficient and the kernel coefficient equal 1.0 and $1/|\mathbf{f}_{\text{CN-LBP}}|$, respectively.
- Random forests (RF): The criterion function is "entropy", and the number of decision trees equals 100.
- K-nearest neighbours (KNN): The number of k-neighbours is 5, and the distance metric is the standard Euclidean metric.

To evaluate the performance, we adopt the micro accuracy scores that counts the total true positives, false negatives, and false positives. Moreover, it is worth noting that CUReT, KTH-TIPS2-b, and Outex-TC-00013 consist of natural texture images, but SIRI-WHU only contains remote sensing images (RSI). All experiment results are averaged over a large number of independent experiments through Python, including NUMPY, OPENCV, NETWORKX, SKIMAGE, etc.

#### B. CN-LBP vs. LBP-based Variants

In Table I, we present the optimal classification accuracies of CN-LBP and LBP-based variants on CUReT, KTH-TIPS2-b, and Outex-TC-00013, respectively, where the best classifier is SVM-L. As shown in Table I, we can see that CN-LBP gives better performance compared with the previous works, where the classification accuracies on three datasets reach 92.95%, 98.78%, and 83.08%, respectively. Especially, except for the local texture information, our proposed texture descriptor contains more detailed image information, including color (bands), structure (gradient), and spatial-relationship (pixel distribution), etc, which better describes the characteristics of texture images. Moreover, with the help of LBP encoding, CN-LBP is rotation invariant and has certain abilities to reduce the impact of illumination.

TABLE I
CLASSIFICATION ACCURACIES (%) OF CN-LBP AND DIFFERENT LBP-BASED VARIANTS ON THREE DATASETS

| Method | CUReT | KTH-TIPS2-b | Outex-TC-00013 |
|---|---|---|---|
| LBP [10] | 91.03 | 89.63 | 77.97 |
| LDP [21] | 85.89 | 82.05 | 76.99 |
| LEP [22] | 77.03 | 76.50 | 73.00 |
| LDN [23] | 84.67 | 81.32 | 74.04 |
| DCP [24] | 87.86 | 77.76 | 70.83 |
| DRLBP [25] | 91.29 | 91.74 | 77.55 |
| CSQP [26] | 89.85 | 82.69 | 76.20 |
| LDTP [27] | 92.25 | 90.97 | 80.85 |
| LCCMSP [28] | 90.92 | 93.51 | 80.78 |
| LNDP [29] | 91.64 | 87.85 | 77.16 |
| CN-LBP | 92.95 | 98.78 | 83.08 |

To improve the robustness of CN-LBP, some dimension reducing (DR) methods are also tested on three datasets, as shown in Fig. 2, including principal components analysis (PCA), linear discriminant analysis (LDA), and chi-square test (Chi2), which can effectively measure the importance of each feature value and remove the null bins of spatial histograms. In this article, the number of features of PCA is jointly described and determine by the curve of cumulative explained variance, where the original length of feature vector is 15426, and the numbers of features of LDA and Chi2 are equal to the number of classes in each dataset. From Table II, we can know that CN-LBP+PCA gives the best performance (92.95%, 99.09%, and 83.08%) and CN-LBP+Chi2 has the worst classification accuracies, and the optimal numbers of principal components are equal to 5490, 770, and 1224, respectively.

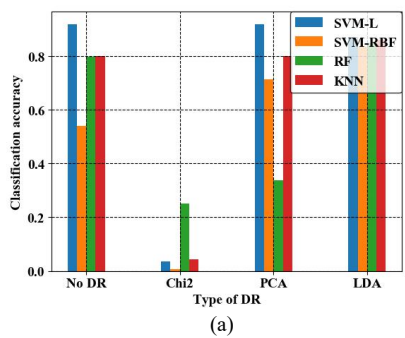

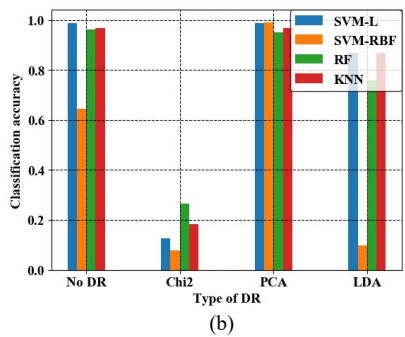

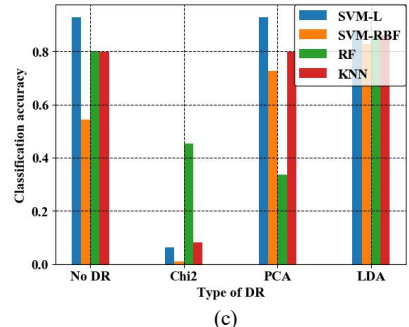

(c)
Fig. 2. Classification accuracies with different classifiers and DR methods. (a) CUReT. (b) KTH-TIPS2-b. (c) Outex-TC-00013.

TABLE II
CLASSIFICATION ACCURACIES (%) OF CN-LBP ASSOCIATED WITH DIFFERENT DR METHODS ON THREE DATASETS

| Method | CUReT | KTH-TIPS2-b | Outex-TC-00013 |
|---|---|---|---|
| CN-LBP+PCA | 92.95 | 99.09 | 83.08 |
| CN-LBP+LDA | 88.03 | 86.97 | 75.98 |
| CN-LBP+Chi2 | 33.60 | 26.36 | 24.27 |

### C. CN-LBP vs. Deep learning-based Methods

In Table III, we present the optimal classification accuracies of CN-LBP and deep learning-based methods on SIRI-WHU, where the best classifier is SVM-L. As listed in [1]-[39], the deep learning-based models include local-global feature bag-of-visual-words (LGFBOVW), deep features and sparse topics (DST), normalized DST (NDST), normalized deep features and 255 stretched sparse topics (ND255ST), and adaptive deep sparse semantic modeling (ADSSM). As illustrated in Table III, we can know that the classification accuracy of CN-LBP reaches 98.06% that is close to ND255ST and ADSSM, and better than LGFBOVW, DST, and NDST. Moreover, from the perspective of computing cost, CN-LBP has higher computing efficiency and can save more operation time compared with deep learning-based models, since CN-LBP does not spend time on training parameters.

TABLE III
CLASSIFICATION ACCURACIES (%) OF CN-LBP AND DIFFERENT DEEP LEARNING-BASED METHODS ON SIRI-WHU

| Method | SIRI-WHU |
|---|---|
| LGFBOVW | 96.96 |
| DST | 91.92 |
| NDST | 95.91 |
| ND255ST | 99.25 |
| ADSSM | 99.75 |
| CN-LBP | 98.06 |

## IV. CONCLUSION

For accurate texture classification, this article proposed a new texture descriptor named CN-LBP by combining the characteristics of CN and LBP. Specifically, we first modeled a TI as directed graphs according to pixel distance, intensity, and gradient. Next, four feature images were generated by CN-based feature measurements. Then, we encoded TIs, GIs, and feature images by ULBP. Finally, CN-LBP was obtained by jointly calculating and concatenating six spatial histograms. The experiment results on four datasets showed that CN-LBP had better classification accuracies compared with other LBP-based variants and deep learning-based methods. Therefore, the proposed texture descriptor could make some contributions to texture classification.